\documentclass[aps,pra,twocolumn,showpacs,amsmath,amssymb]{revtex4}

\usepackage{graphicx}
\usepackage{dcolumn}
\usepackage{bm}
\usepackage{textcomp,mathcomp}

\begin{document}
\title{Effect of magnetization inhomogeneity on magnetic microtraps for atoms}

\author{S. Whitlock}
\author{B. V. Hall}
\email{brhall@swin.edu.au}
\author{T. Roach}
\author{R. Anderson}
\author{P. Hannaford}
\author{A. I. Sidorov}

\affiliation{ ARC Centre of Excellence for Quantum-Atom Optics and
\\Centre for Atom Optics and Ultrafast Spectroscopy, Swinburne
University of Technology, Hawthorn, Victoria 3122, Australia }

\date{\today}

\begin{abstract}
We report on the origin of fragmentation of ultracold atoms observed on a
magnetic film atom chip. Radio frequency spectroscopy and optical imaging of the trapped atoms is used to characterize small
spatial variations of the magnetic field near the film surface. Direct observations indicate the fragmentation
is due to a corrugation of the magnetic potential caused by long range
inhomogeneity in the film magnetization. A model which takes into account
two-dimensional variations of the film magnetization is consistent with the observations.

\end{abstract}

\pacs{03.75.Be, 07.55.Ge, 34.50.Dy, 39.25.+k}%

\maketitle


An atom chip is designed to manipulate magnetically trapped
ultracold atoms near a surface using an arrangement of
microfabricated wires or patterned magnetic materials \cite{Fol02}.
Since the realization of Bose-Einstein condensates (BECs) on atom
chips \cite{han01,ott01}, pioneering experiments have studied single-mode propagation
along waveguides \cite{lea02}, transport and adiabatic
splitting of a BEC \cite{hom05} and recently on-chip atom
interferometry \cite{shi05,sch05}.  Permanent magnets are particularly attractive for
atom chips as they can provide complex magnetic potentials~\cite{sid02} while suppressing current noise that
causes heating and limits the lifetime of trapped atoms near a surface \cite{sin05b}.
To date, permanent magnet atom chips have been developed
with a view to study one-dimensional quantum gases \cite{ven04,llo05,boy06},
decoherence of BEC near surfaces \cite{sin05b,hal06}, hybrid magnetic and optical trapping
configurations \cite{Jaa05}, and self biased fully permanent magnetic potentials \cite{bar05}.
 It has been found, however, that in addition to current noise, atom chips have other limitations, as undesired spatial magnetic field
variations associated with the current-carrying wires or magnetic materials act
to fragment the trapped atoms.

In previous work, fragmentation of atoms trapped near
current-carrying wires was traced to roughness of the wire edges that causes tiny current deviations \cite{wan04,est04}.
This introduces a spatially varying magnetic field component
parallel to the wire which corrugates the bottom of the trap
potential.  While more advanced microfabrication
techniques have been used to produce wires with
extremely straight edges, thereby minimizing fragmentation \cite{Kru04,Pie05}, the first experiments with permanent magnet atom chips have now also indicated
the presence of significant fragmentation \cite{sin05a,hal05,boy06}.
This has motivated further work towards understanding the
mechanisms that cause fragmentation near magnetic materials.

\begin{figure}
\includegraphics{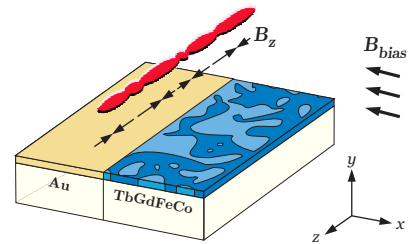}
\caption{\label{fig:schem} (Color online) Schematic diagram of our permanent magnetic
film atom chip.  Inhomogeneity in the film magnetization leads to fragmentation of the trapped cloud of ultracold atoms when positioned near
the surface.}
\end{figure}

In this paper we report on the origin of fragmentation near the surface of a permanent magnetic film atom chip.
To characterize the magnetic field near the film surface we have
developed a technique which combines precision radio frequency
(rf) spectroscopy of trapped atoms with high spatial resolution optical imaging.
This allows sensitive and intrinsically calibrated measurements of
the magnetic field landscape to be made over a large area.
We find the fragmentation originates from long
range inhomogeneity in the film magnetization and has characteristics that differ from those observed for current-carrying
wire atom chips. To account for the observations we have developed a model for the spatial decay of random
magnetic fields from the surface due to inhomogeneity in the film magnetization.

A schematic diagram of a basic magnetic film atom chip is shown in Fig. \ref{fig:schem}.
Our atom chip uses a Tb$_6$Gd$_{10}$Fe$_{80}$Co$_4$ film which exhibits strong perpendicular anisotropy~\cite{hal06}.  The edge
of the $300~$\textmu m thick glass-slide substrate is polished to optical quality prior to film
deposition. Scanning
profilometer measurements on similarly prepared substrates indicate
that the residual edge roughness is less than 50~nm and the top
surface is extremely smooth. The substrate is sputter-coated with a
multilayer magnetic film ($6\times150$~nm TbGdFeCo and
$6\times140$~nm Cr) and a gold overlayer (100~nm), and the film
topology accurately follows that of the polished substrate. The deposited film has been analyzed using a SQUID
magnetometer and a magnetic force microscope
(MFM) and has shown excellent magnetic properties \cite{wan05}. The film is
magnetized perpendicular to the surface by a magnetic field of
1~T and afterwards is magnetically homogeneous within the sensitivity
of the MFM. A second glass slide, coated with a non-magnetic gold
film, completes the reflective atom chip surface (Fig. \ref{fig:schem}). Both substrates
are then epoxied to a $500~$\textmu m thick silver foil
current-carrying structure which is used for loading ultracold atoms into the permanent magnet
microtrap, to provide
weak longitudinal confinement, and as an in-built radio frequency antenna.

\begin{figure}
\includegraphics{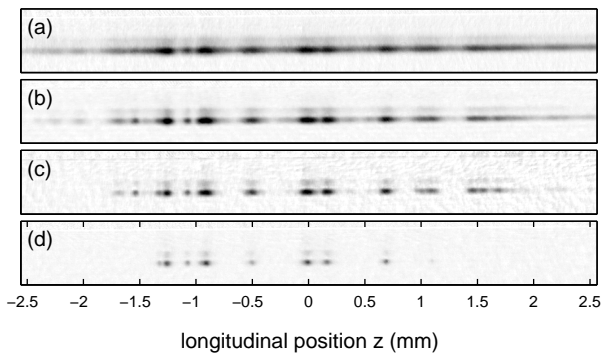}
\caption{\label{fig:density} Absorption images of the atomic density
in the magnetic microtrap located $y_0=67~$\textmu m from the surface.  As the rf
cut-off $\nu_f$ is decreased, the structure of the potential is revealed. (a) $\nu_f$=1238~kHz, (b) $\nu_f$=890~kHz, (c) $\nu_f$=766~kHz, (d) $\nu_f$=695~kHz.}
\end{figure}

At the edge of the perpendicularly magnetized film a field is produced that is
analogous to that of a thin current-carrying wire aligned with the film edge ($I_{eff}=\nobreak0.2~$A) \cite{sid02,hal06}. A magnetic
microtrap is formed by the field from the film, a uniform magnetic
field $B_{bias}$, and two current-carrying end-wires. In the experiment $2\times10^8$ $^{87}$Rb atoms are collected in a mirror magneto-optical
trap located 5~mm from the surface. These atoms are optically pumped
to the $|F=\nobreak2,m_F=\nobreak+2\rangle$ hyperfine state and subsequently
transferred to a magnetic trap formed by a Z-shaped current-carrying wire and $B_{bias}$.
A preliminary rf evaporative cooling stage is used
to reduce the cloud temperature below 5~\textmu K. The remaining
atoms are then transferred to the magnetic film microtrap by
adiabatically reducing the current through the Z-shaped wire to zero. The
final values of $B_{bias}$ vary from 0.2~mT to 0.8~mT, so the
transverse trap frequency varies between $2\pi\times410$~Hz and
$2\pi\times1500$~Hz while the trap position $y_0$ ranges from
200~\textmu m to 50~\textmu m from the surface. The end-wires are
operated at 0.5~A such that the trap depth is $\sim$100~\textmu K and the elongated cloud of atoms
extends 5~mm along the edge of the atom chip to allow measurement of the magnetic potential.

The narrow energy distribution of ultracold atoms is an inherent
advantage when used as a probe of weak potentials. In particular, the
equilibrium distribution of trapped atoms has been used to image
magnetic fields near test wires with high sensitivity and high
spatial resolution \cite{wil05,gun05}.  In parallel, rf
spectroscopy has been used as a precise and powerful method for
investigating the properties of cold atom clouds \cite{mar88,hel92,blo99,gup03,chi04}. In this paper we
use rf spectroscopy of ultracold atoms to accurately profile small magnetic field variations near the
magnetic film surface.

A spatially uniform rf field of frequency $\nu$ is applied perpendicular to the trap
axis to resonantly outcouple atoms to untrapped magnetic states at positions where $g_F\mu_B |B(x,y,z)|=h\nu$.
The rf field is swept using a single frequency
ramp ($\sim$0.2~MHz/s) from 2~MHz to a final cut-off frequency $\nu_f$ ranging between
1.4~MHz and 0.5~MHz.  The Rabi frequency of the rf transition is $2\pi\times0.5~$kHz, high
enough to ensure that atoms with total energy greater than $h\nu$
are removed from the trap with high probability and that regions of
the potential where $g_F\mu_B |B(x,y,z)|\geq h\nu$ thereafter remain
unpopulated. During the early stages of the rf sweep the cloud
undergoes some evaporative cooling as the in-trap collision rate is
high enough to allow rethermalization. At the end of the sweep the resonant frequency approaches that corresponding to the
trap bottom and the cloud becomes significantly truncated by
the rf field.

Immediately after the sweep, $B_{bias}$ is switched off to
accelerate the atoms away from the film surface in the remaining permanent magnetic field gradient.
The longitudinal density distribution is unperturbed during the 1~ms expansion time and is an accurate representation of the
in-trap distribution. A resonant absorption image is then recorded
by a CCD with a spatial resolution of $5$~\textmu m. A series of
absorption images for different values of $\nu_f$ is shown in
Fig.~\ref{fig:density}. Noticeable fragmentation is observed when $\nu_f$
is reduced below 1.3~MHz (Fig.~\ref{fig:density}a). For
$\nu_f\sim0.9$~MHz the density distribution becomes truncated by
the rf field and regions of the atomic density decrease to zero
(Fig.~\ref{fig:density}b). Reducing $\nu_f$ further
results in well separated clumps of atoms which are found only in
the lowest potential wells (Fig.~\ref{fig:density}c, d).

For a quantitative analysis we assume that the full trapping potential
can be expressed in terms of the transverse confinement and the corrugated longitudinal potential
$m_F g_F \mu_B |B_z(z)+B_0|$ where $B_0$ is the value of the uniform offset field~\cite{est04}. The atomic distribution in the trap immediately
after the rf sweep is described by a truncated Boltzmann
distribution \cite{hel92,lui96}. To extract $|B_z(z)+B_0|$, the integrated atomic density
as a function of $\nu_{f}$ can be fit for each position $z$ to the
truncated thermal distribution function,

\begin{equation}
\label{eq:nvsrf}
 n(z,\beta) =n_\infty(z)[\text{erf}(\sqrt{\beta})-2\sqrt{\beta/\pi}\text{e}^{-\beta}(1+2\beta/3)],
\end{equation} where $n_\infty$ is the integrated atom density before truncation and the
spatially dependent truncation parameter $\beta$ is
\begin{equation}
\label{eq:trunc} \beta(z,\nu_f) =(m_F h\nu_f-m_Fg_F\mu_B|B_z(z)+B_0|)/k_BT,
\end{equation}where $T$ is a fit parameter, which characterizes the non-equilibrium
distribution during truncation.  We find however that this model does not satisfactorily reproduce the density distribution in regions of the potential
where the atomic density becomes large; specifically at the bottom of each potential well for lower values of $\nu_f$.  To minimise this effect we section the full data set into 80~$\mu$m spatial regions and fit Eq.~\eqref{eq:nvsrf} using a two-dimensional minimisation algorithm as opposed to fitting for each value of $z$ independently.  This effectively constrains the fitted truncation temperature $T$ to vary smoothly over a range corresponding roughly to the extent of a potential well.  After reconstructing the magnetic field profile the effect of the end-wires is subtracted.
The statistical uncertainty in the measurement is approximately
0.1~\textmu T, which is mainly attributed to fluctuations of external magnetic
fields. With appropriate magnetic shielding the expected sensitivity
of the technique is limited by the power broadened rf
linewidth required to effectively outcouple atoms.



\begin{figure}
\includegraphics{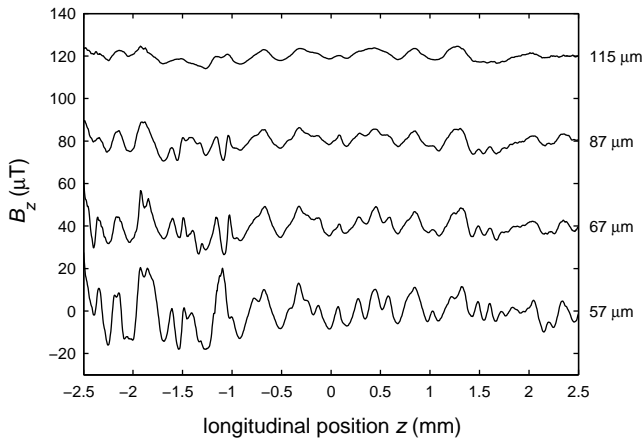}
\caption{\label{fig:bfield} Magnetic field profiles measured using
spatially resolved rf spectroscopy for various distances to the surface. The field offset $B_0$ and the effect of
weak longitudinal confinement are subtracted and each profile has been
offset by 40~\textmu T for clarity.}
\end{figure}

Complete magnetic field profiles are given for several distances
from the film in Fig.~\ref{fig:bfield}.  Firstly, the amplitude and
structure of the corrugated potential are constant from day to day;
however the amplitude increases as the trap is positioned closer to
the surface, with an approximate power law dependence given by
$y^{-a}$, where $a=1.8\pm0.3$. For $y_0>100$~\textmu m the
potential has a characteristic period of about 390~\textmu m,
significantly longer than that commonly observed near electroplated
wires \cite{for02,est04}. Closer to the film, additional
corrugations appear with a characteristic period of about 170~\textmu m.
These $y$-dependent characteristics of the potential have allowed
time-dependent manipulation of BECs at particularly interesting
regions of the disordered potential \cite{hal06b}.


The figure of merit for corrugation observed above the TbGdFeCo film is $B_{z,rms}/B(y)\sim4\times10^{-2}$ which compares poorly with the current generation of lithographically fabricated wire atom chips $B_{z,rms}/B(y)\sim1\times10^{-4}$~\cite{Kru04}, highlighting the need for further work to improve magnetic materials used in these applications.

\begin{figure}
\includegraphics[width=7.5cm]{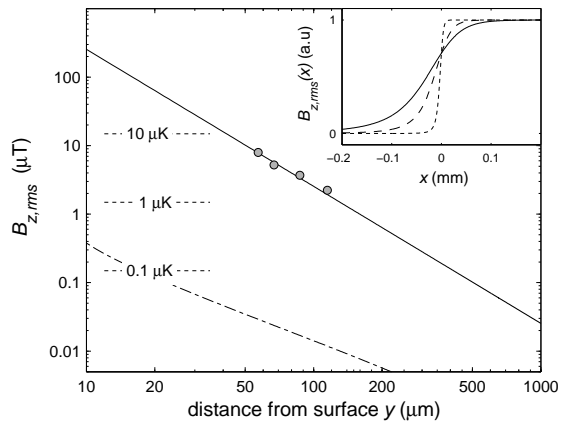}
\caption{\label{fig:model} Behavior of the magnetic field roughness $B_{z,rms}$
as a function of distance from the film surface. The dashed line
corresponds to the situation of a homogeneous film with a
fluctuating edge as measured by a profilometer. The solid line is a fit to the data using Eq.
\ref{eq:modelfinal} which corresponds to white noise magnetization
variations within the body of the film with a characteristic feature
size $d=5$~\textmu m and $\Delta M/M_s\simeq$0.3.  The inset shows the predicted $x$ dependence of the field roughness away from the film edge for distances of $y=10~\mu$m (dotted), $y=50~\mu$m (dashed) and $y=100~\mu$m (solid line).}
\end{figure}

We have found that the amplitude of the corrugated potential is not consistent with magnetostatic calculations based on fluctuations of
the film edge (Fig. \ref{fig:model}), suggesting that the cause of
fragmentation may be spatial variations of the magnetization within
the body of the film. To investigate this, a second bias field is
applied in the $y$ direction to bring the trap closer to the surface
while keeping a constant distance from the edge of the magnetic
film.  A cloud of atoms positioned 50~\textmu m from the surface of
the magnetic film and 100~\textmu m from the edge is significantly
fragmented, while a cloud positioned 50~\textmu m from the
non-magnetic gold film does not exhibit atomic density variations.
This observation confirms that the magnetic field variations
originate from the magnetic material itself and not from imperfections along the edge of the film.

Our model describes the effect of two-dimensional spatial variations
in the perpendicular magnetization $M_y(x,z)$ of the film.
Inhomogeneity leads to the appearance of a random magnetic field
above the surface, of which we are most interested in the magnetic field
component $B_z(x,y,z)$ that corrugates the bottom of the trapping
potential. Using a standard approach incorporating the
two-dimensional Fourier transform of the random magnetization
$N(k_x,k_z)$ and the magnetic scalar potential we obtain an
expression for the $B_z$ component of the corrugated magnetic field.
In the case of a  magnetic film occupying a half-plane with the edge
at $x=0$ (Fig. \ref{fig:schem}) and arbitrary magnetization noise
we have, for heights greater than the film thickness,
\begin{eqnarray}
B_z = i2\pi\mu_0 y\delta \int_{-\infty}^{\infty}
\int_{-\infty}^{\infty} dk_x dk_z k_z^2 N(k_x,k_z){\rm
e}^{i2\pi (k_x x+k_z z)} \nonumber\\
\times \int_{-x}^{\infty} dx'{\rm e}^{i2\pi k_x x'} \frac{K_1 (2\pi
k_z \sqrt{x'^2+y^2})}{\sqrt{x'^2+y^2}},
\end{eqnarray}
where $\delta$ is the film thickness and $K_1$ is the modified
Bessel function of the second kind. In general this expression can
be used to describe any planar pattern of elements that can be represented as a linear combination of
step functions.

In the case of white noise fluctuations in the magnetization,
$|N(k_x,k_z)| = const$, the \textit{rms} value of the magnetic field
roughness can be evaluated analytically. We perform ensemble
averaging and find
\begin{equation}
B_{z,rms}= \sqrt{\frac{3}{\pi}}\frac{\mu_0 d \delta\Delta M}{16
y^2}\sqrt{1 + \frac{15}{8}\alpha - \frac{5}{4}\alpha^3 +
\frac{3}{8}\alpha^5}, \label{eq:modelfinal}
\end{equation}
where $\alpha = x / \sqrt{x^2 + y^2}$, $\Delta M$ is the
\textit{rms} magnetization inhomogeneity and $d$ is the
characteristic feature size of the domain structure. For $x=0$ the
model predicts that the corrugated magnetic field component decays
with a $y^{-2}$ dependence, consistent with our
experimental result. This behavior
can also be compared with the more rapid decay $(\propto y^{-2.5})$
expected for white noise fluctuations of the edge of
current-carrying wires \cite{wan04,est04}. Film edge fluctuations
are expected to produce corrugations three orders of magnitude
smaller than that observed in the experiments (Fig. \ref{fig:model}). The model also
predicts the fast decay of the corrugated magnetic field away from the film
edge for $x<0$~(Fig.~\ref{fig:model}-inset).

In Fig.~\ref{fig:model} the experimental results are compared with
the model with relevant energy scales indicated by dotted lines. The
characteristic feature size and distribution function of the domain
structure has been inferred from MFM measurements of a demagnetized
TbGdFeCo film and is found to have close to white noise
characteristics with $d\approx5~$\textmu m.  The best fit of
Eq.~\eqref{eq:modelfinal} to the data is found for $\Delta M/M_s\simeq$
0.3 where $M_s$ is the saturation magnetization of the film.  If the
 inhomogeneity is assumed to originate from reversal
of a small number of magnetic domains ($M_y(x,z)=\pm M_s$) then we conclude that
the mean magnetization of the film is greater than $0.9M_s$.

The TbGdFeCo magnetic film was originally chosen for its desirable
magnetic properties including a large coercivity ($\mu_0 H_c=\nobreak0.32$~T)
and a high Curie temperature ($T_c\sim$300$^\circ$C) \cite{wan05}.
We attribute the observed inhomogeneity to deterioration of the magnetic film
experienced during the vacuum bake-out (140~$^\circ$C over 4~days) despite the relatively high
Curie temperature of our film.  This conclusion is consistent with
reports of reduced perpendicular anisotropy found for similar films
after annealing at temperatures above 100$^\circ$C
\cite{lub85,wan90} and with our own measurements on similar films.

In conclusion, trapped ultracold atoms are very sensitive to
small magnetic field variations found near the surface of the
permanent magnetic film.  These variations corrugate the
longitudinal trapping potential and result in fragmentation of atomic
density. We have developed the technique of spatially resolved rf spectroscopy
as a powerful method for accurately mapping small magnetic field variations near the surface of the magnetic film.  A simple
model accounts for spatial inhomogeneity of the film magnetization
and agrees well with the observations.  The development of new
permanent magnet atom chips will require additional research aimed
at further optimizing the quality of magnetic films.

\begin{acknowledgments}
We would like to thank J. Wang for the deposition of the magnetic
film. T. Roach acknowledges the Donors of the American Chemical Society Petroleum Research Fund for support during this research. This project is supported by the ARC Centre of Excellence for
Quantum-Atom Optics and a Swinburne University Strategic Initiative
grant.
\end{acknowledgments}

\bibliography{swhitlock_pra}

\begin{thebibliography}{34}
\expandafter\ifx\csname natexlab\endcsname\relax\def\natexlab#1{#1}\fi
\expandafter\ifx\csname bibnamefont\endcsname\relax
  \def\bibnamefont#1{#1}\fi
\expandafter\ifx\csname bibfnamefont\endcsname\relax
  \def\bibfnamefont#1{#1}\fi
\expandafter\ifx\csname citenamefont\endcsname\relax
  \def\citenamefont#1{#1}\fi
\expandafter\ifx\csname url\endcsname\relax
  \def\url#1{\texttt{#1}}\fi
\expandafter\ifx\csname urlprefix\endcsname\relax\def\urlprefix{URL }\fi
\providecommand{\bibinfo}[2]{#2}
\providecommand{\eprint}[2][]{\url{#2}}

\bibitem[{\citenamefont{Folman et~al.}(2002)\citenamefont{Folman, Kr{\"{u}}ger,
  Schmiedmayer, Denschlag, and Henkel}}]{Fol02}
\bibinfo{author}{\bibfnamefont{R.}~\bibnamefont{Folman}},
  \bibinfo{author}{\bibfnamefont{P.}~\bibnamefont{Kr{\"{u}}ger}},
  \bibinfo{author}{\bibfnamefont{J.}~\bibnamefont{Schmiedmayer}},
  \bibinfo{author}{\bibfnamefont{J.}~\bibnamefont{Denschlag}},
  \bibnamefont{and} \bibinfo{author}{\bibfnamefont{C.}~\bibnamefont{Henkel}},
  \bibinfo{journal}{Adv. At. Mol. Opt. Phys.} \textbf{\bibinfo{volume}{48}},
  \bibinfo{pages}{263} (\bibinfo{year}{2002}).

\bibitem[{\citenamefont{H{\"{a}}nsel et~al.}(2001)\citenamefont{H{\"{a}}nsel,
  Hommelhoff, H{\"{a}}nsch, and Reichel}}]{han01}
\bibinfo{author}{\bibfnamefont{W.}~\bibnamefont{H{\"{a}}nsel}},
  \bibinfo{author}{\bibfnamefont{P.}~\bibnamefont{Hommelhoff}},
  \bibinfo{author}{\bibfnamefont{T.~W.} \bibnamefont{H{\"{a}}nsch}},
  \bibnamefont{and} \bibinfo{author}{\bibfnamefont{J.}~\bibnamefont{Reichel}},
  \bibinfo{journal}{Nature (London)} \textbf{\bibinfo{volume}{413}},
  \bibinfo{pages}{498} (\bibinfo{year}{2001}).

\bibitem[{\citenamefont{Ott et~al.}(2001)\citenamefont{Ott, Fort{\'{a}}gh,
  Schlotterbeck, Grossmann, and Zimmermann}}]{ott01}
\bibinfo{author}{\bibfnamefont{H.}~\bibnamefont{Ott}},
  \bibinfo{author}{\bibfnamefont{J.}~\bibnamefont{Fort{\'{a}}gh}},
  \bibinfo{author}{\bibfnamefont{G.}~\bibnamefont{Schlotterbeck}},
  \bibinfo{author}{\bibfnamefont{A.}~\bibnamefont{Grossmann}},
  \bibnamefont{and}
  \bibinfo{author}{\bibfnamefont{C.}~\bibnamefont{Zimmermann}},
  \bibinfo{journal}{Phys.\ Rev.\ Lett.} \textbf{\bibinfo{volume}{87}},
  \bibinfo{pages}{230401} (\bibinfo{year}{2001}).

\bibitem[{\citenamefont{Leanhardt et~al.}(2002)\citenamefont{Leanhardt,
  Chikkatur, Kielpinski, Shin, Gustavson, Ketterle, and Pritchard}}]{lea02}
\bibinfo{author}{\bibfnamefont{A.~E.} \bibnamefont{Leanhardt}},
  \bibinfo{author}{\bibfnamefont{A.~P.} \bibnamefont{Chikkatur}},
  \bibinfo{author}{\bibfnamefont{D.}~\bibnamefont{Kielpinski}},
  \bibinfo{author}{\bibfnamefont{Y.}~\bibnamefont{Shin}},
  \bibinfo{author}{\bibfnamefont{T.~L.} \bibnamefont{Gustavson}},
  \bibinfo{author}{\bibfnamefont{W.}~\bibnamefont{Ketterle}}, \bibnamefont{and}
  \bibinfo{author}{\bibfnamefont{D.~E.} \bibnamefont{Pritchard}},
  \bibinfo{journal}{Phys.\ Rev.\ Lett.} \textbf{\bibinfo{volume}{89}},
  \bibinfo{pages}{040401} (\bibinfo{year}{2002}).

\bibitem[{\citenamefont{Hommelhoff et~al.}(2005)\citenamefont{Hommelhoff,
  H{\"{a}}nsel, Steinmetz, H{\"{a}}nsch, and Reichel}}]{hom05}
\bibinfo{author}{\bibfnamefont{P.}~\bibnamefont{Hommelhoff}},
  \bibinfo{author}{\bibfnamefont{W.}~\bibnamefont{H{\"{a}}nsel}},
  \bibinfo{author}{\bibfnamefont{T.}~\bibnamefont{Steinmetz}},
  \bibinfo{author}{\bibfnamefont{T.~W.} \bibnamefont{H{\"{a}}nsch}},
  \bibnamefont{and} \bibinfo{author}{\bibfnamefont{J.}~\bibnamefont{Reichel}},
  \bibinfo{journal}{New J. Phys.} \textbf{\bibinfo{volume}{7}},
  \bibinfo{pages}{3} (\bibinfo{year}{2005}).

\bibitem[{\citenamefont{Shin et~al.}(2005)\citenamefont{Shin, Sanner, Jo,
  Pasquini, Saba, Ketterle, Pritchard, Vengalattore, and Prentiss}}]{shi05}
\bibinfo{author}{\bibfnamefont{Y.}~\bibnamefont{Shin}},
  \bibinfo{author}{\bibfnamefont{C.}~\bibnamefont{Sanner}},
  \bibinfo{author}{\bibfnamefont{G.~B.} \bibnamefont{Jo}},
  \bibinfo{author}{\bibfnamefont{T.~A.} \bibnamefont{Pasquini}},
  \bibinfo{author}{\bibfnamefont{M.}~\bibnamefont{Saba}},
  \bibinfo{author}{\bibfnamefont{W.}~\bibnamefont{Ketterle}},
  \bibinfo{author}{\bibfnamefont{D.~E.} \bibnamefont{Pritchard}},
  \bibinfo{author}{\bibfnamefont{M.}~\bibnamefont{Vengalattore}},
  \bibnamefont{and} \bibinfo{author}{\bibfnamefont{M.}~\bibnamefont{Prentiss}},
  \bibinfo{journal}{Phys.\ Rev.\ A} \textbf{\bibinfo{volume}{72}},
  \bibinfo{pages}{021604(R)} (\bibinfo{year}{2005}).

\bibitem[{\citenamefont{Schumm et~al.}(2005)\citenamefont{Schumm, Hofferberth,
  Andersson, Wildermuth, Groth, Bar-Joseph, Schmiedmayer, and
  Kr{\"{u}}ger}}]{sch05}
\bibinfo{author}{\bibfnamefont{T.}~\bibnamefont{Schumm}},
  \bibinfo{author}{\bibfnamefont{S.}~\bibnamefont{Hofferberth}},
  \bibinfo{author}{\bibfnamefont{L.~M.} \bibnamefont{Andersson}},
  \bibinfo{author}{\bibfnamefont{S.}~\bibnamefont{Wildermuth}},
  \bibinfo{author}{\bibfnamefont{S.}~\bibnamefont{Groth}},
  \bibinfo{author}{\bibfnamefont{I.}~\bibnamefont{Bar-Joseph}},
  \bibinfo{author}{\bibfnamefont{J.}~\bibnamefont{Schmiedmayer}},
  \bibnamefont{and}
  \bibinfo{author}{\bibfnamefont{P.}~\bibnamefont{Kr{\"{u}}ger}},
  \bibinfo{journal}{Nature (London)} \textbf{\bibinfo{volume}{1}},
  \bibinfo{pages}{57} (\bibinfo{year}{2005}).

\bibitem[{\citenamefont{Sidorov et~al.}(2002)\citenamefont{Sidorov, McLean,
  Scharnberg, Gough, Davis, Sexton, Opat, and Hannaford}}]{sid02}
\bibinfo{author}{\bibfnamefont{A.~I.} \bibnamefont{Sidorov}},
  \bibinfo{author}{\bibfnamefont{R.~J.} \bibnamefont{McLean}},
  \bibinfo{author}{\bibfnamefont{F.}~\bibnamefont{Scharnberg}},
  \bibinfo{author}{\bibfnamefont{D.~S.} \bibnamefont{Gough}},
  \bibinfo{author}{\bibfnamefont{T.~J.} \bibnamefont{Davis}},
  \bibinfo{author}{\bibfnamefont{B.~A.} \bibnamefont{Sexton}},
  \bibinfo{author}{\bibfnamefont{G.~I.} \bibnamefont{Opat}}, \bibnamefont{and}
  \bibinfo{author}{\bibfnamefont{P.}~\bibnamefont{Hannaford}},
  \bibinfo{journal}{Acta Phys. Pol. B} \textbf{\bibinfo{volume}{33}},
  \bibinfo{pages}{2137} (\bibinfo{year}{2002}).

\bibitem[{\citenamefont{Sinclair
  et~al.}(2005{\natexlab{a}})\citenamefont{Sinclair, Curtis, Garcia, Retter,
  Hall, Eriksson, Sauer, and Hinds}}]{sin05b}
\bibinfo{author}{\bibfnamefont{C.~D.~J.} \bibnamefont{Sinclair}},
  \bibinfo{author}{\bibfnamefont{E.~A.} \bibnamefont{Curtis}},
  \bibinfo{author}{\bibfnamefont{I.~L.} \bibnamefont{Garcia}},
  \bibinfo{author}{\bibfnamefont{J.~A.} \bibnamefont{Retter}},
  \bibinfo{author}{\bibfnamefont{B.~V.} \bibnamefont{Hall}},
  \bibinfo{author}{\bibfnamefont{S.}~\bibnamefont{Eriksson}},
  \bibinfo{author}{\bibfnamefont{B.~E.} \bibnamefont{Sauer}}, \bibnamefont{and}
  \bibinfo{author}{\bibfnamefont{E.~A.} \bibnamefont{Hinds}},
  \bibinfo{journal}{Phys.\ Rev.\ A} \textbf{\bibinfo{volume}{72}},
  \bibinfo{pages}{031603(R)} (\bibinfo{year}{2005}{\natexlab{a}}).

\bibitem[{\citenamefont{Vengalattore et~al.}(2004)\citenamefont{Vengalattore,
  Conroy, Rooijakkers, and Prentiss}}]{ven04}
\bibinfo{author}{\bibfnamefont{M.}~\bibnamefont{Vengalattore}},
  \bibinfo{author}{\bibfnamefont{R.~S.} \bibnamefont{Conroy}},
  \bibinfo{author}{\bibfnamefont{W.}~\bibnamefont{Rooijakkers}},
  \bibnamefont{and} \bibinfo{author}{\bibfnamefont{M.}~\bibnamefont{Prentiss}},
  \bibinfo{journal}{J. Appl. Phys.} \textbf{\bibinfo{volume}{95}},
  \bibinfo{pages}{4404} (\bibinfo{year}{2004}).

\bibitem[{\citenamefont{Llorente-Garcia
  et~al.}(2005)\citenamefont{Llorente-Garcia, Sinclair, Curtis, Eriksson,
  Sauer, and Hinds}}]{llo05}
\bibinfo{author}{\bibfnamefont{I.}~\bibnamefont{Llorente-Garcia}},
  \bibinfo{author}{\bibfnamefont{C.~D.~J.} \bibnamefont{Sinclair}},
  \bibinfo{author}{\bibfnamefont{E.~A.} \bibnamefont{Curtis}},
  \bibinfo{author}{\bibfnamefont{S.}~\bibnamefont{Eriksson}},
  \bibinfo{author}{\bibfnamefont{B.~E.} \bibnamefont{Sauer}}, \bibnamefont{and}
  \bibinfo{author}{\bibfnamefont{E.~A.} \bibnamefont{Hinds}},
  \bibinfo{journal}{J. Phys.: Conf. Ser.} \textbf{\bibinfo{volume}{19}},
  \bibinfo{pages}{70} (\bibinfo{year}{2005}).

\bibitem[{\citenamefont{Boyd et~al.}(2006)\citenamefont{Boyd, Streed, Medley,
  Campbell, Mun, Ketterle, and Pritchard}}]{boy06}
\bibinfo{author}{\bibfnamefont{M.}~\bibnamefont{Boyd}},
  \bibinfo{author}{\bibfnamefont{E.~W.} \bibnamefont{Streed}},
  \bibinfo{author}{\bibfnamefont{P.}~\bibnamefont{Medley}},
  \bibinfo{author}{\bibfnamefont{G.~K.} \bibnamefont{Campbell}},
  \bibinfo{author}{\bibfnamefont{J.}~\bibnamefont{Mun}},
  \bibinfo{author}{\bibfnamefont{W.}~\bibnamefont{Ketterle}}, \bibnamefont{and}
  \bibinfo{author}{\bibfnamefont{D.~E.} \bibnamefont{Pritchard}},
  \bibinfo{journal}{cond-mat/0608370}  (\bibinfo{year}{2006}).

\bibitem[{\citenamefont{Hall et~al.}(2006{\natexlab{a}})\citenamefont{Hall,
  Whitlock, Scharnberg, Hannaford, and Sidorov}}]{hal06}
\bibinfo{author}{\bibfnamefont{B.~V.} \bibnamefont{Hall}},
  \bibinfo{author}{\bibfnamefont{S.}~\bibnamefont{Whitlock}},
  \bibinfo{author}{\bibfnamefont{F.}~\bibnamefont{Scharnberg}},
  \bibinfo{author}{\bibfnamefont{P.}~\bibnamefont{Hannaford}},
  \bibnamefont{and} \bibinfo{author}{\bibfnamefont{A.}~\bibnamefont{Sidorov}},
  \bibinfo{journal}{J. Phys. B: At. Mol. Opt. Phys.}
  \textbf{\bibinfo{volume}{39}}, \bibinfo{pages}{27}
  (\bibinfo{year}{2006}{\natexlab{a}}).

\bibitem[{\citenamefont{Jaakkola et~al.}(2005)\citenamefont{Jaakkola,
  Shevchenko, Lindfors, Hautakorpi, Il'yashenko, Johansen, and
  Kaivola}}]{Jaa05}
\bibinfo{author}{\bibfnamefont{A.}~\bibnamefont{Jaakkola}},
  \bibinfo{author}{\bibfnamefont{A.}~\bibnamefont{Shevchenko}},
  \bibinfo{author}{\bibfnamefont{K.}~\bibnamefont{Lindfors}},
  \bibinfo{author}{\bibfnamefont{M.}~\bibnamefont{Hautakorpi}},
  \bibinfo{author}{\bibfnamefont{E.}~\bibnamefont{Il'yashenko}},
  \bibinfo{author}{\bibfnamefont{T.~H.} \bibnamefont{Johansen}},
  \bibnamefont{and} \bibinfo{author}{\bibfnamefont{M.}~\bibnamefont{Kaivola}},
  \bibinfo{journal}{Eur. Phys. J. D} \textbf{\bibinfo{volume}{35}},
  \bibinfo{pages}{81} (\bibinfo{year}{2005}).

\bibitem[{\citenamefont{Barb et~al.}(2005)\citenamefont{Barb, Gerritsma, Xing,
  Goedkoop, and Spreeuw}}]{bar05}
\bibinfo{author}{\bibfnamefont{I.}~\bibnamefont{Barb}},
  \bibinfo{author}{\bibfnamefont{R.}~\bibnamefont{Gerritsma}},
  \bibinfo{author}{\bibfnamefont{Y.~T.} \bibnamefont{Xing}},
  \bibinfo{author}{\bibfnamefont{J.~B.} \bibnamefont{Goedkoop}},
  \bibnamefont{and} \bibinfo{author}{\bibfnamefont{R.~J.~C.}
  \bibnamefont{Spreeuw}}, \bibinfo{journal}{Eur. Phys. J. D}
  \textbf{\bibinfo{volume}{35}}, \bibinfo{pages}{75} (\bibinfo{year}{2005}).

\bibitem[{\citenamefont{Wang et~al.}(2004)\citenamefont{Wang, Lukin, and
  Demler}}]{wan04}
\bibinfo{author}{\bibfnamefont{D.~W.} \bibnamefont{Wang}},
  \bibinfo{author}{\bibfnamefont{M.~D.} \bibnamefont{Lukin}}, \bibnamefont{and}
  \bibinfo{author}{\bibfnamefont{E.}~\bibnamefont{Demler}},
  \bibinfo{journal}{Phys.\ Rev.\ Lett.} \textbf{\bibinfo{volume}{92}},
  \bibinfo{pages}{076802} (\bibinfo{year}{2004}).

\bibitem[{\citenamefont{Est{\`{e}}ve et~al.}(2004)\citenamefont{Est{\`{e}}ve,
  Aussibal, Schumm, Figl, Mailly, Bouchoule, Westbrook, and Aspect}}]{est04}
\bibinfo{author}{\bibfnamefont{J.}~\bibnamefont{Est{\`{e}}ve}},
  \bibinfo{author}{\bibfnamefont{C.}~\bibnamefont{Aussibal}},
  \bibinfo{author}{\bibfnamefont{T.}~\bibnamefont{Schumm}},
  \bibinfo{author}{\bibfnamefont{C.}~\bibnamefont{Figl}},
  \bibinfo{author}{\bibfnamefont{D.}~\bibnamefont{Mailly}},
  \bibinfo{author}{\bibfnamefont{I.}~\bibnamefont{Bouchoule}},
  \bibinfo{author}{\bibfnamefont{C.~I.} \bibnamefont{Westbrook}},
  \bibnamefont{and} \bibinfo{author}{\bibfnamefont{A.}~\bibnamefont{Aspect}},
  \bibinfo{journal}{Phys.\ Rev.\ A} \textbf{\bibinfo{volume}{70}},
  \bibinfo{pages}{043629} (\bibinfo{year}{2004}).

\bibitem[{\citenamefont{Kr{\"{u}}ger et~al.}(2004)\citenamefont{Kr{\"{u}}ger,
  Andersson, Wildermuth, Hofferberth, Haller, Aigner, Groth, Bar-Joseph, and
  Schmiedmayer}}]{Kru04}
\bibinfo{author}{\bibfnamefont{P.}~\bibnamefont{Kr{\"{u}}ger}},
  \bibinfo{author}{\bibfnamefont{L.~M.} \bibnamefont{Andersson}},
  \bibinfo{author}{\bibfnamefont{S.}~\bibnamefont{Wildermuth}},
  \bibinfo{author}{\bibfnamefont{S.}~\bibnamefont{Hofferberth}},
  \bibinfo{author}{\bibfnamefont{E.}~\bibnamefont{Haller}},
  \bibinfo{author}{\bibfnamefont{S.}~\bibnamefont{Aigner}},
  \bibinfo{author}{\bibfnamefont{S.}~\bibnamefont{Groth}},
  \bibinfo{author}{\bibfnamefont{I.}~\bibnamefont{Bar-Joseph}},
  \bibnamefont{and}
  \bibinfo{author}{\bibfnamefont{J.}~\bibnamefont{Schmiedmayer}},
  \bibinfo{journal}{cond-mat/0504686}  (\bibinfo{year}{2004}).

\bibitem[{\citenamefont{Pietra et~al.}(2005)\citenamefont{Pietra, Aigner, vom
  Hagen, Lezec, and Schmiedmayer}}]{Pie05}
\bibinfo{author}{\bibfnamefont{L.~D.} \bibnamefont{Pietra}},
  \bibinfo{author}{\bibfnamefont{S.}~\bibnamefont{Aigner}},
  \bibinfo{author}{\bibfnamefont{C.}~\bibnamefont{vom Hagen}},
  \bibinfo{author}{\bibfnamefont{H.~J.} \bibnamefont{Lezec}}, \bibnamefont{and}
  \bibinfo{author}{\bibfnamefont{J.}~\bibnamefont{Schmiedmayer}},
  \bibinfo{journal}{J. Phys.: Conf. Ser.} \textbf{\bibinfo{volume}{19}},
  \bibinfo{pages}{30} (\bibinfo{year}{2005}).

\bibitem[{\citenamefont{Sinclair
  et~al.}(2005{\natexlab{b}})\citenamefont{Sinclair, Retter, Curtis, Hall,
  Llorente-Garcia, Eriksson, Sauer, and Hinds}}]{sin05a}
\bibinfo{author}{\bibfnamefont{C.~D.~J.} \bibnamefont{Sinclair}},
  \bibinfo{author}{\bibfnamefont{J.~A.} \bibnamefont{Retter}},
  \bibinfo{author}{\bibfnamefont{E.~A.} \bibnamefont{Curtis}},
  \bibinfo{author}{\bibfnamefont{B.~V.} \bibnamefont{Hall}},
  \bibinfo{author}{\bibfnamefont{I.}~\bibnamefont{Llorente-Garcia}},
  \bibinfo{author}{\bibfnamefont{S.}~\bibnamefont{Eriksson}},
  \bibinfo{author}{\bibfnamefont{B.~E.} \bibnamefont{Sauer}}, \bibnamefont{and}
  \bibinfo{author}{\bibfnamefont{E.~A.} \bibnamefont{Hinds}},
  \bibinfo{journal}{Eur. Phys. J. D} \textbf{\bibinfo{volume}{35}},
  \bibinfo{pages}{105} (\bibinfo{year}{2005}{\natexlab{b}}).

\bibitem[{\citenamefont{Hall et~al.}(2005)\citenamefont{Hall, Whitlock,
  Scharnberg, Hannaford, and Sidorov}}]{hal05}
\bibinfo{author}{\bibfnamefont{B.~V.} \bibnamefont{Hall}},
  \bibinfo{author}{\bibfnamefont{S.}~\bibnamefont{Whitlock}},
  \bibinfo{author}{\bibfnamefont{F.}~\bibnamefont{Scharnberg}},
  \bibinfo{author}{\bibfnamefont{P.}~\bibnamefont{Hannaford}},
  \bibnamefont{and} \bibinfo{author}{\bibfnamefont{A.}~\bibnamefont{Sidorov}},
  in \emph{\bibinfo{booktitle}{Laser Spectroscopy XVII}}, edited by
  \bibinfo{editor}{\bibfnamefont{E.~A.} \bibnamefont{Hinds}},
  \bibinfo{editor}{\bibfnamefont{A.}~\bibnamefont{Ferguson}}, \bibnamefont{and}
  \bibinfo{editor}{\bibfnamefont{E.}~\bibnamefont{Riis}}
  (\bibinfo{publisher}{World Scientific, Singapore}, \bibinfo{year}{2005}).

\bibitem[{\citenamefont{Wang et~al.}(2005)\citenamefont{Wang, Whitlock,
  Scharnberg, Gough, Sidorov, McLean, and Hannaford}}]{wan05}
\bibinfo{author}{\bibfnamefont{J.~Y.} \bibnamefont{Wang}},
  \bibinfo{author}{\bibfnamefont{S.}~\bibnamefont{Whitlock}},
  \bibinfo{author}{\bibfnamefont{F.}~\bibnamefont{Scharnberg}},
  \bibinfo{author}{\bibfnamefont{D.}~\bibnamefont{Gough}},
  \bibinfo{author}{\bibfnamefont{A.~I.} \bibnamefont{Sidorov}},
  \bibinfo{author}{\bibfnamefont{R.~J.} \bibnamefont{McLean}},
  \bibnamefont{and}
  \bibinfo{author}{\bibfnamefont{P.}~\bibnamefont{Hannaford}},
  \bibinfo{journal}{J. Phys. D: Appl. Phys.} \textbf{\bibinfo{volume}{38}},
  \bibinfo{pages}{4015} (\bibinfo{year}{2005}).

\bibitem[{\citenamefont{Wildermuth et~al.}(2005)\citenamefont{Wildermuth,
  Hofferberth, Lesanovsky, Haller, Andersson, Groth, Bar-Joseph, Kr{\"{u}}ger,
  and Schmiedmayer}}]{wil05}
\bibinfo{author}{\bibfnamefont{S.}~\bibnamefont{Wildermuth}},
  \bibinfo{author}{\bibfnamefont{S.}~\bibnamefont{Hofferberth}},
  \bibinfo{author}{\bibfnamefont{I.}~\bibnamefont{Lesanovsky}},
  \bibinfo{author}{\bibfnamefont{E.}~\bibnamefont{Haller}},
  \bibinfo{author}{\bibfnamefont{L.~M.} \bibnamefont{Andersson}},
  \bibinfo{author}{\bibfnamefont{S.}~\bibnamefont{Groth}},
  \bibinfo{author}{\bibfnamefont{I.}~\bibnamefont{Bar-Joseph}},
  \bibinfo{author}{\bibfnamefont{P.}~\bibnamefont{Kr{\"{u}}ger}},
  \bibnamefont{and}
  \bibinfo{author}{\bibfnamefont{J.}~\bibnamefont{Schmiedmayer}},
  \bibinfo{journal}{Nature (London)} \textbf{\bibinfo{volume}{435}},
  \bibinfo{pages}{440} (\bibinfo{year}{2005}).

\bibitem[{\citenamefont{G{\"{u}}nther et~al.}(2005)\citenamefont{G{\"{u}}nther,
  Kemmler, Kraft, Vale, Zimmermann, and Fort{\'{a}}gh}}]{gun05}
\bibinfo{author}{\bibfnamefont{A.}~\bibnamefont{G{\"{u}}nther}},
  \bibinfo{author}{\bibfnamefont{M.}~\bibnamefont{Kemmler}},
  \bibinfo{author}{\bibfnamefont{S.}~\bibnamefont{Kraft}},
  \bibinfo{author}{\bibfnamefont{C.~J.} \bibnamefont{Vale}},
  \bibinfo{author}{\bibfnamefont{C.}~\bibnamefont{Zimmermann}},
  \bibnamefont{and}
  \bibinfo{author}{\bibfnamefont{J.}~\bibnamefont{Fort{\'{a}}gh}},
  \bibinfo{journal}{Phys.\ Rev.\ A} \textbf{\bibinfo{volume}{71}},
  \bibinfo{pages}{063619} (\bibinfo{year}{2005}).

\bibitem[{\citenamefont{Martin et~al.}(1988)\citenamefont{Martin, Helmerson,
  Bagnato, Lafyatis, and Pritchard}}]{mar88}
\bibinfo{author}{\bibfnamefont{A.~G.} \bibnamefont{Martin}},
  \bibinfo{author}{\bibfnamefont{K.}~\bibnamefont{Helmerson}},
  \bibinfo{author}{\bibfnamefont{V.~S.} \bibnamefont{Bagnato}},
  \bibinfo{author}{\bibfnamefont{G.~P.} \bibnamefont{Lafyatis}},
  \bibnamefont{and} \bibinfo{author}{\bibfnamefont{D.~E.}
  \bibnamefont{Pritchard}}, \bibinfo{journal}{Phys.\ Rev.\ Lett.}
  \textbf{\bibinfo{volume}{61}}, \bibinfo{pages}{2431} (\bibinfo{year}{1988}).

\bibitem[{\citenamefont{Helmerson et~al.}(1992)\citenamefont{Helmerson, Martin,
  and Pritchard}}]{hel92}
\bibinfo{author}{\bibfnamefont{K.}~\bibnamefont{Helmerson}},
  \bibinfo{author}{\bibfnamefont{A.~G.} \bibnamefont{Martin}},
  \bibnamefont{and} \bibinfo{author}{\bibfnamefont{D.~E.}
  \bibnamefont{Pritchard}}, \bibinfo{journal}{J. Opt. Soc. Am. B}
  \textbf{\bibinfo{volume}{9}}, \bibinfo{pages}{483} (\bibinfo{year}{1992}).

\bibitem[{\citenamefont{Bloch et~al.}(1999)\citenamefont{Bloch, H{\"{a}}nsch,
  and Esslinger}}]{blo99}
\bibinfo{author}{\bibfnamefont{I.}~\bibnamefont{Bloch}},
  \bibinfo{author}{\bibfnamefont{T.~W.} \bibnamefont{H{\"{a}}nsch}},
  \bibnamefont{and}
  \bibinfo{author}{\bibfnamefont{T.}~\bibnamefont{Esslinger}},
  \bibinfo{journal}{Phys.\ Rev.\ Lett.} \textbf{\bibinfo{volume}{82}},
  \bibinfo{pages}{3008} (\bibinfo{year}{1999}).

\bibitem[{\citenamefont{Gupta et~al.}(2003)\citenamefont{Gupta, Hadzibabic,
  Zwierlein, Stan, Dieckmann, Schunck, van Kempen, Verhaar, and
  Ketterle}}]{gup03}
\bibinfo{author}{\bibfnamefont{S.}~\bibnamefont{Gupta}},
  \bibinfo{author}{\bibfnamefont{Z.}~\bibnamefont{Hadzibabic}},
  \bibinfo{author}{\bibfnamefont{M.~W.} \bibnamefont{Zwierlein}},
  \bibinfo{author}{\bibfnamefont{C.~A.} \bibnamefont{Stan}},
  \bibinfo{author}{\bibfnamefont{K.}~\bibnamefont{Dieckmann}},
  \bibinfo{author}{\bibfnamefont{C.~H.} \bibnamefont{Schunck}},
  \bibinfo{author}{\bibfnamefont{E.~G.~M.} \bibnamefont{van Kempen}},
  \bibinfo{author}{\bibfnamefont{B.~J.} \bibnamefont{Verhaar}},
  \bibnamefont{and} \bibinfo{author}{\bibfnamefont{W.}~\bibnamefont{Ketterle}},
  \bibinfo{journal}{Science} \textbf{\bibinfo{volume}{300}},
  \bibinfo{pages}{1723} (\bibinfo{year}{2003}).

\bibitem[{\citenamefont{Chin et~al.}(2004)\citenamefont{Chin, Bartenstein,
  Altmeyer, Riedl, Jochim, Denschlag, and Grimm}}]{chi04}
\bibinfo{author}{\bibfnamefont{C.}~\bibnamefont{Chin}},
  \bibinfo{author}{\bibfnamefont{M.}~\bibnamefont{Bartenstein}},
  \bibinfo{author}{\bibfnamefont{A.}~\bibnamefont{Altmeyer}},
  \bibinfo{author}{\bibfnamefont{S.}~\bibnamefont{Riedl}},
  \bibinfo{author}{\bibfnamefont{S.}~\bibnamefont{Jochim}},
  \bibinfo{author}{\bibfnamefont{J.~H.} \bibnamefont{Denschlag}},
  \bibnamefont{and} \bibinfo{author}{\bibfnamefont{R.}~\bibnamefont{Grimm}},
  \bibinfo{journal}{Science} \textbf{\bibinfo{volume}{305}},
  \bibinfo{pages}{1128} (\bibinfo{year}{2004}).

\bibitem[{\citenamefont{Luiten et~al.}(1996)\citenamefont{Luiten, Reynolds, and
  Walraven}}]{lui96}
\bibinfo{author}{\bibfnamefont{O.~J.} \bibnamefont{Luiten}},
  \bibinfo{author}{\bibfnamefont{M.~W.} \bibnamefont{Reynolds}},
  \bibnamefont{and} \bibinfo{author}{\bibfnamefont{J.~T.~M.}
  \bibnamefont{Walraven}}, \bibinfo{journal}{Phys.\ Rev.\ A}
  \textbf{\bibinfo{volume}{53}}, \bibinfo{pages}{381} (\bibinfo{year}{1996}).

\bibitem[{\citenamefont{Fort{\'{a}}gh et~al.}(2002)\citenamefont{Fort{\'{a}}gh,
  Ott, Kraft, G{\"{u}}nther, and Zimmermann}}]{for02}
\bibinfo{author}{\bibfnamefont{J.}~\bibnamefont{Fort{\'{a}}gh}},
  \bibinfo{author}{\bibfnamefont{H.}~\bibnamefont{Ott}},
  \bibinfo{author}{\bibfnamefont{S.}~\bibnamefont{Kraft}},
  \bibinfo{author}{\bibfnamefont{A.}~\bibnamefont{G{\"{u}}nther}},
  \bibnamefont{and}
  \bibinfo{author}{\bibfnamefont{C.}~\bibnamefont{Zimmermann}},
  \bibinfo{journal}{Phys.\ Rev.\ A} \textbf{\bibinfo{volume}{66}},
  \bibinfo{pages}{041604(R)} (\bibinfo{year}{2002}).

\bibitem[{\citenamefont{Hall et~al.}(2006{\natexlab{b}})\citenamefont{Hall,
  Whitlock, Anderson, Hannaford, and Sidorov}}]{hal06b}
\bibinfo{author}{\bibfnamefont{B.~V.} \bibnamefont{Hall}},
  \bibinfo{author}{\bibfnamefont{S.}~\bibnamefont{Whitlock}},
  \bibinfo{author}{\bibfnamefont{R.}~\bibnamefont{Anderson}},
  \bibinfo{author}{\bibfnamefont{P.}~\bibnamefont{Hannaford}},
  \bibnamefont{and} \bibinfo{author}{\bibfnamefont{A.}~\bibnamefont{Sidorov}},
  \bibinfo{journal}{cond-mat/0609014}  (\bibinfo{year}{2006}{\natexlab{b}}).

\bibitem[{\citenamefont{Luborsky}(1985)}]{lub85}
\bibinfo{author}{\bibfnamefont{F.~E.} \bibnamefont{Luborsky}},
  \bibinfo{journal}{J. Appl. Phys.} \textbf{\bibinfo{volume}{57}},
  \bibinfo{pages}{3592} (\bibinfo{year}{1985}).

\bibitem[{\citenamefont{Wang and Leng}(1990)}]{wan90}
\bibinfo{author}{\bibfnamefont{Y.~J.} \bibnamefont{Wang}} \bibnamefont{and}
  \bibinfo{author}{\bibfnamefont{Q.~W.} \bibnamefont{Leng}},
  \bibinfo{journal}{Phys.\ Rev.\ B} \textbf{\bibinfo{volume}{41}},
  \bibinfo{pages}{651} (\bibinfo{year}{1990}).

\end{thebibliography}

\end{document}